%% file: fm19.tex
\documentclass[envcountsame,envcountsect]{llncs}

\usepackage{helvet}
\usepackage{courier}
\usepackage{amsmath}
\usepackage{amssymb}
\usepackage{array}
\usepackage{algorithm}
\usepackage{algorithmicx}
\usepackage{subcaption}
\usepackage{nicematrix}
\usepackage{multirow}

\usepackage[noend]{algpseudocode}
\usepackage{tikz}
\usetikzlibrary{arrows,automata,calc}
\input{macros}

\title{Defensive Design of Saturating Counters Based on Differential Privacy}

\begin{document}

\author{
Depeng Liu\inst{1,3} \and
Lutan Zhao\inst{2,3} \and
Pengfei Yang\inst{1,3} \and
Bow-Yaw Wang \inst{4} \and
Rui Hou\inst{2,3} \and
Lijun Zhang\inst{1,3} \and
Naijun Zhan\inst{1,3}}

\authorrunning{D. Liu et al.}

\institute{State Key Laboratory of Computer Science, Institute of Software, Chinese Academy of Sciences, Beijing, China
\email{\{liudp,yangpf,zhanglj,zhannj\}@ios.ac.cn}\\ \and
State Key Laboratory of Information Security, Institute of Information Engineering, Chinese Academy of Sciences, Beijing, China
\email{\{zhaolutan,hourui\}@iie.ac.cn}\\ \and
University of Chinese Academy of Sciences, Beijing, China \and
Institute of Information Science, Academia Sinica, Taiwan
\email{bywang@iis.sinica.edu.tw}\\
}

\maketitle

\begin{abstract}
The saturating counter is the basic module of the dynamic branch predictor, which involves the core technique to improve instruction level parallelism performance in modern processors. However, most studies focus on the performance improvement and hardware consumption of saturating counters, while ignoring the security problems they may cause. In this paper, we creatively propose to study and design saturating counters from the defense perspective of differential privacy, so that attackers cannot distinguish the states that saturating counters are in and further infer sensitive information. To obtain theoretical guarantees, we use Markov chain to formalize the attack algorithm applied to the saturating counter, investigate into the optimal attack strategy and calculate the probability of successful attack. Furthermore, we find that the attacker is able to accurately guess the branch execution of the victim's process in the existing saturating counters. To avoid this, we design a new probabilistic saturating counter, which generalizes the existing conventional and probabilistic saturating counters. The guarantee of differential privacy is applied to deduce parameters of the new saturating counters so that the security requirement can be satisfied. We also theoretically calculate the misprediction rate when the saturating counter reaches the steady state. The experimental results on testing programs show that the calculated theoretical results agree with the experimental performances. Compared with the existing conventional and probabilistic saturating counters, when the parameters of our designed models are selected appropriately, the new saturating counters can not only ensure similar operational performance, but also establish strict security guarantee.
\end{abstract}

\input{introduction}

\section{Differential Privacy on Probabilistic Saturating Counters}

The success attack of Algorithm~\ref{algorithm:cut-off} lies in the deterministic behaviors of the saturating counters. That is, given a state and an input, the post state is determined surely. To make efficient defense against such attack, ~\cite{JCST2021zhao} designs a probabilistic saturating counter (PSC). It introduces a probabilistic threshold in the counter. When a transition is made, a random number is generated and compared with the threshold. If the generated number is bigger, the transition is carried on; Otherwise the counter remains in its current state. The model becomes a Markov chain and is depicted in Fig.~\ref{fig:psc2}. For each state $i$ where $i$ is the binary code of the state, $P_{iT}$ is the statistical probability of executing the target branch with taken direction and $P_{iN}$ is the probability with not-taken direction and $P_{iT}+P_{iN}=1$. The probability threshold $m$ is a number in $[0,1]$ and $n=1-m$. For instance, in state $3$ (ST), there is a probability of $mP_{3T}+n$ to remain in state $3$ and of $mP_{3N}$ to get to state $2$ (WT).

In the PSC setting, the attacker cannot precisely predict the action, because the number of steps it takes to get to the cut-off point is not deterministic any more. The experiments in \cite{JCST2021zhao} shows that the experimental attack success rate and the performance in the simulation environment under different values of the parameter $m$. However, only experimental results cannot make up for the theoretical analysis of the optimal attack strategy on the PSC, as well as theoretical guarantees of the misprediction rate, which is important in the performance evaluation. Meanwhile, we note that differential privacy aims at guaranteeing similar output probability distributions on neighboring inputs so that an attacker cannot distinguish. In a similar way, we would like to make defense on PSCs and we hope that the attacker cannot distinguish the victim's execution. Thus, we apply differential privacy to PSCs and establish theoretical guarantee of the defense. What's more, we analysis the optimal attack strategy of Algorithm~\ref{algorithm:cut-off} and discover the scenario where the attacker can always infer the victim's execution successfully. To avoid this, we design a new PSC model and calculate the misprediction rate of the new model. Above all, we define differential privacy on PSCs:

\begin{definition}
A PSC satisfies $(\epsilon,\delta)$-differential privacy, if and only if for every output $c$ of Algorithm~\ref{algorithm:cut-off}, the probability of getting $c$ are bounded by the following inequalities:
\begin{align}
\Pr[\textit{out}=c|v=\text{T}] &\leq e^\epsilon \Pr[\textit{out}=c|v=\text{NT}] + \delta \label{ineq:1},\\
\Pr[\textit{out}=c|v=\text{NT}] &\leq e^\epsilon \Pr[\textit{out}=c|v=\text{T}] + \delta \label{ineq:2},
\end{align}
where T/NT represents that the victim executes the branch with taken/not-taken direction.
\label{def:pscdp}
\end{definition}

Note that it is quite straightforward to apply differential privacy here. Algorithm~\ref{algorithm:cut-off} corresponds to the privacy mechanism in the original definition and it is requested that the probability distributions of the outputs on neighbors are mathematically similar. Here we only have one pair of neighbors, which are represented by the victim executing the branch with taken and not-taken direction.  When the attacker performs the attacks, he/she cannot distinguish whether the victim's branch is taken or not-taken.
Thus we protect the ``privacy'' of victim's execution. Besides, the definition is general for all forms of PSCs when Algorithm~\ref{algorithm:cut-off} is performed.

\begin{figure}
    \centering
     \resizebox{0.6\textwidth}{!}{
    \begin{tikzpicture}[->,>=stealth',shorten >=1pt,auto,node
      distance=2cm,node/.style={circle,draw,inner sep=0pt,minimum size=35pt}]
      \node[node, fill = red!50, label = {}]
      (ST) at (-2, 2) {ST(3)};
      \node[node, fill = red!50, label = {}]
      (WT) at (2, 2) {WT(2)};
      \node[node, fill = blue!50, label = {}]
      (WN) at (-2, -2) {WN(1)};
      \node[node, fill = blue!50, label = {}]
      (SN) at (2, -2) {SN(0)};

      \path
      (ST) edge [bend left] node [above] {$mP_{3\mathrm N}$} (WT)
      (ST) edge [loop left] node [] {$mP_{3\mathrm T} + n$} (ST)
      (WT) edge [bend left] node [below] {$mP_{2\mathrm T}$} (ST)
      (WT) edge [loop right] node [right] {$n$} (WT)

      (SN) edge [bend left] node [below] {$mP_{0\mathrm T}$} (WN)
      (SN) edge [loop right] node [] {$mP_{0\mathrm N} + n$} (SN)
      (WN) edge [bend left] node [above] {$mP_{1\mathrm N}$} (SN)
      (WN) edge [loop left] node [left] {$n$} (WN)

      (WT) edge [bend left] node [right] {$mP_{2\mathrm N}$} (SN)
      (WN) edge [bend left] node [left] {$mP_{1\mathrm T}$} (ST)
      ;
    \end{tikzpicture}}
    \caption{A 2-bit probabilistic saturating counter.}
    \label{fig:psc2}
\end{figure}

\section{Model Attack Algorithm and Resolve Optimal Strategy}\label{sec:counter-strategy}
The definition of differential privacy on PSCs enlightens us to observe the probability of getting every output under different execution of victim's branch and we try to figure out when given an output $c$, what is the attacker's strategy to determine whether the branch is taken or not-taken. The experiments in ~\cite{JCST2021zhao} fix a value of probability threshold $m$ and simulate the victim's behavior for plenty of times. The attacker will keep record of the number of different outputs when the branch is taken or not-taken, then tries to guess the victim's behavior with the direction where the number shows more often. The drawback of this method is to perform the algorithm for a large amount of times and cannot obtain general conclusions when $m$ is arbitrary. We will use the Markov chain to model the algorithm and analysis the best attack strategy by the model.

During the three phases when Algorithm~\ref{algorithm:cut-off} is performed on PSCs, the directions of the branches executed are different, as well as probabilities:
\begin{itemize}
    \item In Phase $1$, the branch is always taken. Thus for every state $i$, $P_{i\mathrm T} = 1$ and $P_{i\mathrm N} = 0$.
    \item In Phase $2$, the branch is executed once depending on the victim's thread. So $P_{i\mathrm T} = 1$ or $P_{i\mathrm T} = 0$.
    \item In Phase $3$, the branch is always not-taken. Thus for every state $i$, $P_{i\mathrm T} = 0$ and $P_{i\mathrm N} = 1$.
\end{itemize}

When $m=0$, every state in the PSC will always remain and no transition will occur. As a result, the PSC's prediction becomes static and it will always predict taken (or not-taken). The misprediction rate is $50\%$ on average. We assume $m\neq 0$. Then the attacker can send enough prime signals to make PSC in state ST. Then we use the modeling language PRISM~\cite{KNP:11:PRISM} to build the model of Algorithm~\ref{algorithm:cut-off} and consider the situation when the victim executes the branch with taken and not-taken direction. The Markov chain generated is depicted in Fig.~\ref{fig:2mcs}. Note that the attacker and victim thread's behaviors both exist in the model. According to victim's execution in Phase 2, the model can be divided into 2 sub models. The first transition of each sub model represents the victim's execution. State ST' and ST are the same state in the original PSC, but with the victim finishing executing its branch or not. So in state ST', the victim has already executed its branch. After the first transition, the PSC will keep reading NT and making transitions until reaching state SN.

\begin{figure}
\begin{subfigure}{0.49\textwidth}
     \centering
       \resizebox{\textwidth}{!}{
    \begin{tikzpicture}[->,>=stealth',shorten >=1pt,auto,node
      distance=2cm,node/.style={circle,draw,inner sep=0pt,minimum size=35pt}]
      \draw[-stealth'] (-2.9,1.5) -- (-2,1.5);
      \node[node, fill = red!50, label = {}]
      (ST) at (-1.5, 1.5) {ST(3)};
      \node[node, fill = red!50, label = {}]
      (ST2) at (1.5, 1.5) {ST'(3)};
      \node[node, fill = blue!50, label = {}]
      (SN) at (-1.5, -1.5) {SN(0)};
      \node[node, fill = red!50, label = {}]
      (WT) at (1.5, -1.5) {WT(2)};

      \path
      (ST) edge [left] node [above] {$1$} (ST2)
      (ST2) edge [left] node [right] {$m$} (WT)
      (ST2) edge [loop right] node [right] {$1-m$} (ST2)

      (WT) edge [left] node [below] {$m$} (SN)
      (WT) edge [loop right] node [] {$1-m$} (WT)
      (SN) edge [loop left] node [left] {$1$} (SN)
      ;
    \end{tikzpicture}
    }
\caption{}
    \label{fig:first2}
\end{subfigure}
\begin{subfigure}{0.49\textwidth}
     \centering
       \resizebox{\textwidth}{!}{
    \begin{tikzpicture}[->,>=stealth',shorten >=1pt,auto,node
      distance=2cm,node/.style={circle,draw,inner sep=0pt,minimum size=35pt}]
      \draw[-stealth'] (-2.9,1.5) -- (-2,1.5);
      \node[node, fill = red!50, label = {}]
      (ST) at (-1.5, 1.5) {ST(3)};
      \node[node, fill = red!50, label = {}]
      (ST2) at (1.5, 1.5) {ST'(3)};
      \node[node, fill = blue!50, label = {}]
      (SN) at (-1.5, -1.5) {SN(0)};
      \node[node, fill = red!50, label = {}]
      (WT) at (1.5, -1.5) {WT(2)};

      \path
      (ST) edge [left] node [above] {$m$} (WT)
      (ST) edge [left] node [above] {$1-m$} (ST2)
      (ST2) edge [left] node [right] {$m$} (WT)
      (ST2) edge [loop right] node [right] {$1-m$} (ST2)

      (WT) edge [left] node [below] {$m$} (SN)
      (WT) edge [loop right] node [] {$1-m$} (WT)
      (SN) edge [loop left] node [left] {$1$} (SN)
      ;
    \end{tikzpicture}
    }
    \caption{}
    \label{fig:second2}
\end{subfigure}
\caption{The MC for the attack algorithm of finding the cut-off point when (a) the victim takes the branch, (b) the victim does not take the branch.}
\label{fig:2mcs}
\end{figure}

When $m=1$, PSC degenerates into a deterministic saturating counter.
It can also be seen from Fig. ~\ref{fig:2mcs} that if the victim executes the branch with taken, the PSC must take $3$ transitions to reach state SN, and if the victim executes the branch with not-taken, the PSC must take $2$ transitions to reach state SN, including transitions made by the victim thread. Therefore, the attacker can always predict the execution of the relevant branches according to the different steps to state SN in Phase 3 . When $m \neq 1$, the probability uncertainty makes the number of transitions to state SN uncertain, which unables the attacker to make a direct inference. Therefore, the attacker needs an attack strategy such that, when the number of steps $c$ is observed, whether the victim's branch is taken or not should be determined. Since we model the attack algorithm into a Markov chain, the number of steps $c$ in our model is equal to the number of transitions taken to reach state SN for the first time minus $1$. Our goal becomes to calculate the probability of first reaching SN under different numbers of transitions and compare the probability when the victim's branch is taken or not respectively. Then the attacker's attack strategy is to guess the execution of the victim's branch which has a larger probability. The intuition is similar to how to guess the side of flipping an irregular coin. If the probabilities of head and tail are known,  guessing the side with a larger probability will make the expectation of successfully guessing maximal. In our case, the probability of each side is a function related to the number of transitions, which needs to be calculated.
\begin{figure}
\begin{subfigure}{0.49\textwidth}
     \centering
       \resizebox{\textwidth}{!}{
    \begin{tikzpicture}[->,>=stealth',shorten >=1pt,auto,node
      distance=2cm,node/.style={circle,draw,inner sep=0pt,minimum size=35pt}]
      \node[node, fill = red!50, label = {}]
      (ST) at (-2, 2) {ST(3)};
      \node[node, fill = red!50, label = {}]
      (ST2) at (2, 2) {ST'(3)};
      \node[node, fill = blue!50, label = {}]
      (SN) at (-2, -2) {SN(0)};
      \node[node, fill = red!50, label = {}]
      (WT) at (2, -2) {WT(2)};
      \node[node, fill = yellow!50, label = {}]
      (S) at (-3, 0) {$S$};

      \path
      (ST) edge [left] node [above] {$1$} (ST2)
      (ST2) edge [left] node [right] {$m$} (WT)
      (ST2) edge [loop right] node [right] {$1-m$} (ST2)

      (WT) edge [left] node [below] {$m$} (SN)
      (WT) edge [loop right] node [] {$1-m$} (WT)
      (SN) edge [left] node [left] {$1$} (S)
      (S) edge [loop right] node [] {1} (S)
      ;
    \end{tikzpicture}
    }
\caption{}
    \label{fig:abs1}
\end{subfigure}
\hfill
\begin{subfigure}{0.49\textwidth}
     \centering
       \resizebox{\textwidth}{!}{
    \begin{tikzpicture}[->,>=stealth',shorten >=1pt,auto,node
      distance=2cm,node/.style={circle,draw,inner sep=0pt,minimum size=35pt}]
      \node[node, fill = red!50, label = {}]
      (ST) at (-2, 2) {ST(3)};
      \node[node, fill = red!50, label = {}]
      (ST2) at (2, 2) {ST'(3)};
      \node[node, fill = blue!50, label = {}]
      (SN) at (-2, -2) {SN(0)};
      \node[node, fill = red!50, label = {}]
      (WT) at (2, -2) {WT(2)};
      \node[node, fill = yellow!50, label = {}]
      (S) at (-3, 0) {$S$};

      \path
      (ST) edge [left] node [above] {$m$} (WT)
      (ST) edge [left] node [above] {$1-m$} (ST2)
      (ST2) edge [left] node [right] {$m$} (WT)
      (ST2) edge [loop right] node [right] {$1-m$} (ST2)

      (WT) edge [left] node [below] {$m$} (SN)
      (WT) edge [loop right] node [] {$1-m$} (WT)
      (SN) edge [left] node [left] {$1$} (S)
      (S) edge [loop right] node [] {1} (S)
      ;
    \end{tikzpicture}
    }
    \caption{}
    \label{fig:abs2}
\end{subfigure}
\hfill
\caption{The MC model after introducing an absorbing state $S$, when (a) the victim takes the branch, (b) the victim does not take the branch.}
\label{fig:absorbing}
\end{figure}

Here we make use of the transition matrix of the Markov chain to calculate the probability of arriving state SN for the first time. In Fig.~\ref{fig:2mcs}, we introduce a new absorbing state $S$ and redirect the original transition from state SN to $S$ with a probability of $1$. In the new model, when state SN is reached and passed, the PSC will remain in state $S$. Therefore, calculating the probability of $N$ steps of transitions to reach SN for the first time in the original model is equal to calculating the probability of reaching state SN after $n$ steps in the new model. Formally, let $M_T$ and $M_N$ represent the transition matrix of the new model, respectively. Let $x,y$ be a row vector of dimension $1\times 5$, with support only for state ST(3) in $x$ and support only for SN(0) in $y$. Then, $x\cdot M_T$ and $x\cdot M_N$ represent the probability distribution after the victim executes the branch with taken/not-taken direction, and $(M_T)^c$ and $(M_N)^c$ represent the $c$ steps of transitions, respectively. Therefore, the attack strategy of the attacker is to compare the probability of reaching SN, and select the victim's branch execution with the higher probability. Specifically, we use the mathematical tool SageMath~\cite{zimmermann:18:SageMath}, which is specialized in symbolic algebra calculation, to calculate the probabilities and solve the following inequality,
\begin{align}
x\cdot M_T \cdot (M_T)^C \cdot y^\intercal >
x\cdot M_N \cdot (M_N)^C \cdot y^\intercal.
~\label{ineq:4}
\end{align}
where $y^\intercal$ denotes the transpose of $y$.
\begin{figure}
    \centering
    \includegraphics[width = 0.7\textwidth]{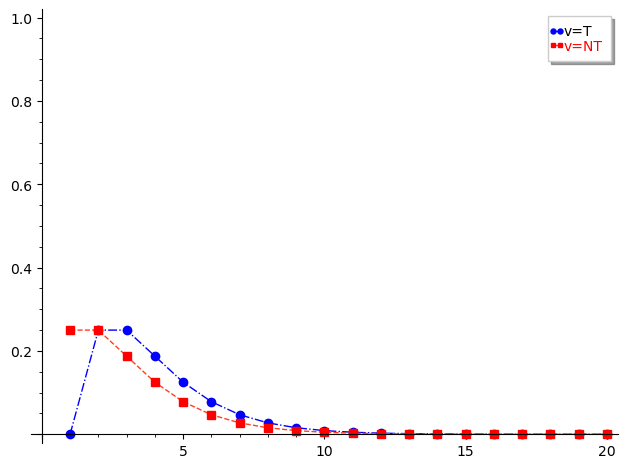}
    \caption{The line chart for the probability of observing the output $c$ when the victim takes the branch ($v=$T) or not ($v=$NT) on a 2-bit PSC when $m=0.5$}
    \label{fig:tnt}
\end{figure}
We obtain a concise solution $c>1/m$, which means when Algorithm
\ref{algorithm:cut-off} outputs a $c$, if $c > 1/m$, then the attacker guesses that the victim executes the branch with taken direction; otherwise, the attacker guesses that the victim executes the branch with not-taken direction. This is the attacker's optimal attack strategy. Next we calculate the probability that an attacker guesses correctly. The left expression of Inequality ~(\ref{ineq:4}) is actually a conditional probability $\Pr[\textit{out}=c|v=\mathrm T]$. Namely, the probability of observing $c$ when victim $v$ executes the branch with taken direction. Similarly the right expression is $\Pr[\textit{out}=c|v=\mathrm{NT}]$, corresponding to when victim $v$ executes the branch with not-taken direction. Note that these two probabilities are exactly the conditional probabilities that need to ensure differential privacy on PSCs in Definition~\ref{def:pscdp}, which will be calculated later. When $c>1/m$, the attacker guesses that the victim execute the branch with taken. The probability of correct guessing, i.e., successful attack, is
\begin{equation}
    \Pr[v=\mathrm T|\textit{out}=c] = \frac{\Pr[v=\mathrm T, \textit{out}=c]}{\Pr[\textit{out}=c]} = \frac{\Pr[\textit{out}=c|v=\mathrm T]\cdot \Pr[v=\mathrm T]}{\Pr[\textit{out}=c]},
    \label{ineq:5}
\end{equation}
where Bayesian rules are applied. Similarly, the probability of wrong guessing is
\begin{equation}
    \Pr[v=\mathrm{NT}|\textit{out}=c] = \frac{\Pr[v=\mathrm{NT}, \textit{out}=c]}{\Pr[\textit{out}=c]} = \frac{\Pr[\textit{out}=c|v=\mathrm{NT}]\cdot \Pr[v=\mathrm{NT}]}{\Pr[\textit{out}=c]},
    \label{ineq:6}
\end{equation}
Assume that the victim's execution branch is uniformly random, thus $\Pr[v=\mathrm{T}]=\Pr[v=\mathrm{NT}]=0.5$. Then with Expression~(\ref{ineq:5})~(\ref{ineq:6}) and $\Pr[v=\mathrm T|\textit{out}=c]+\Pr[v=\mathrm{NT}|\textit{out}=c]=1$, we have
\begin{equation}
    \Pr[v=\mathrm T|\textit{out}=c] =  \frac{\Pr[\textit{out}=c|v=\mathrm T]}{\Pr[\textit{out}=c|v=\mathrm T]+\Pr[\textit{out}=c|v=\mathrm{NT}]},
    \label{ineq:7}
\end{equation}
Similarly, when $c\leq 1/m$, the probability of successful attack is
\begin{equation}
    \Pr[v=\mathrm{NT}|\textit{out}=c] =  \frac{\Pr[\textit{out}=c|v=\mathrm{NT}]}{\Pr[\textit{out}=c|v=\mathrm T]+\Pr[\textit{out}=c|v=\mathrm{NT}]}.
    \label{ineq:8}
\end{equation}

\begin{figure}
    \centering
    \includegraphics[width = 0.7\textwidth]{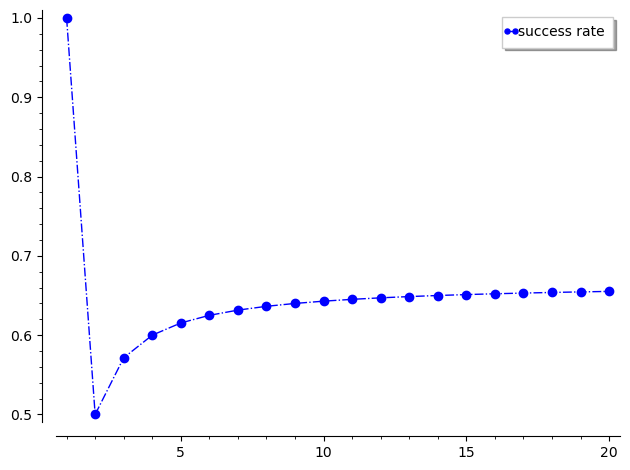}
    \caption{The line chart for the probability of successful attack on a 2-bit PSC when $m=0.5$. }
    \label{fig:rate}
\end{figure}

It is not hard to see the correlation between the probability of successful attack and parameters of differential privacy on PSCs. For the above expressions, if $(\epsilon,0)$-difference privacy on PSCs is guaranteed, the probability of successful attack can be limited within $\frac{e^\epsilon}{1+e^\epsilon}$.
With a fixed $m$, the probability distribution of the observed steps of $C$ when the victim executes the branch with each direction can be computed. Fig. ~\ref{fig:tnt} shows the probability distribution when $m=0.5$. For example, when the victim takes the branch ($v=$T), the probability of observing $c=2$ is $0.25$, which is equal to the probability observed when the victim does not take the branch ($v=$NT). We remark that the graph is basically consistent with the trend of $m=0.5$ in~\cite{JCST2021zhao}, which was obtained by repeated experiments. Differently, our results can be extended to other values of $m$. We can also depict the probability of successful attack of Algorithm ~\ref{algorithm:cut-off} with the change of step $c$, as shown in Fig.~\ref{fig:rate}. In this figure, the probability of successful attack is $1$ when $c=1$. It happens because when the victim takes the branch, it cannot go through $1$ step to reach state SN. Therefore, whenever $c=1$ appears, the attacker will guess correctly that the victim does not take the branch, which can also be observed from the probability distribution of Fig.~\ref{fig:tnt}. When $c=2$, the probability of getting this observation is $0.25$ in both cases, so the probability of successful attack is $0.5$. When $c>2 = 1/m$, the attacker guesses that the victim has a higher probability of executing the branch.

Even if with different values of $m$, the probability of successful attack is $1$ whenever $c=1$ is observed. This is determined by the structure of the PSC itself: If the victim takes the branch, the PSC will remain in state ST and it takes at least $2$ steps to reach state SN, which is not the case when the victim does not take the branch. To avoid this scenario, we design a more secure saturating counter.

\section{New Model and Deduce Privacy Paramters}\label{sec:counter-parameters}

\begin{figure}
 \centering
  \resizebox{\textwidth}{!}{
    \begin{tikzpicture}[->,>=stealth',shorten >=1pt,auto,node
      distance=2cm,node/.style={circle,draw,inner sep=0pt,minimum size=35pt}]
      \node[node, fill = red!50, label = {}]
      (ST) at (-2, 2) {ST(3)};
      \node[node, fill = red!50, label = {}]
      (WT) at (2, 2) {WT(2)};
      \node[node, fill = blue!50, label = {}]
      (WN) at (-2, -2) {WN(1)};
      \node[node, fill = blue!50, label = {}]
      (SN) at (2, -2) {SN(0)};

      \path
      (ST) edge [bend left] node [above] {$m[(1-p)P_{3\mathrm N}+pP_{3\mathrm T}]$} (WT)
      (ST) edge [loop left] node [] {$m[(1-p)P_{3\mathrm T}+pP_{3\mathrm N}] + n$} (ST)
      (WT) edge [bend left] node [below] {$mP_{2\mathrm T}$} (ST)
      (WT) edge [loop right] node [right] {$n$} (WT)

      (SN) edge [bend left] node [below] {$m[(1-p)P_{0\mathrm T}+pP_{0\mathrm N}]$} (WN)
      (SN) edge [loop right] node [] {$m[(1-p)P_{0\mathrm N}+pP_{0\mathrm T}] +n$} (SN)
      (WN) edge [bend left] node [above] {$mP_{1\mathrm N}$} (SN)
      (WN) edge [loop left] node [left] {$n$} (WN)

      (WT) edge [bend left] node [right] {$mP_{2\mathrm N}$} (SN)
      (WN) edge [bend left] node [left] {$mP_{1\mathrm T}$} (ST)
      ;
    \end{tikzpicture}}
    \caption{The MC for the newly designed PSC.}
    \label{figure:nmc}
\end{figure}

As is analyzed before, it is still possible for an attacker to accurately infer the branch execution of the victim process because there exist deterministic transitions in the model. For instance, reading T in state ST remains in state ST. We design a new model based on the PSCs and introduce more probabilities, as shown in Fig.~\ref{figure:nmc}. We retain the original probability update mechanism. That is, the original transition will be updated with a probability $m$. At the same time, a new probability parameter $0\leq p \leq 1$ is introduced in state ST and SN: if reading T in state ST, there is still a probability of $p$ going to state WT, and a probability of $1-p$ to state ST; If reading NT in state ST, there is  a probability of $p$ staying in state ST, and of $1-p$ to state WT. For state SN, a similar way is applied and is necessary on this state. Otherwise, the attacker can change the direction of prime and probe vector in Algorithm~\ref{algorithm:cut-off} to attack and has the same effect as in the previous section. From the perspective of cost and implementation, PSC still uses the probabilistic update mechanism, but only with different update probability for state ST and SN. For state ST, when T is read, there is a probability of $1-mp$ staying in ST, of $mp$ going to state WT; When NT is read, there is a probability of $1-m+mp$ staying in ST and $m-mp$ to state WT. Therefore, whether NT or T is read, as long as $p\neq 0$, there is a probability to stay in state ST or go to state WT. A similar analysis holds for state SN. It can be found that our newly designed PSC generalizes the previous saturating counters: $m=1$, $p=0$, the PSC becomes the conventional saturating counter in Fig.~\ref{fig:sc1}; when $p=0$, PSC becomes the probabilistic saturating counters in~\cite{JCST2021zhao} (Fig.~\ref{fig:psc2}, which we will refer to as original PSCs in the following).

\begin{figure}
\begin{subfigure}{0.49\textwidth}
     \centering
       \resizebox{\textwidth}{!}{
    \begin{tikzpicture}[->,>=stealth',shorten >=1pt,auto,node
      distance=2cm,node/.style={circle,draw,inner sep=0pt,minimum size=35pt}]
      \node[node, fill = red!50, label = {}]
      (ST) at (-2, 2) {ST(3)};
      \node[node, fill = red!50, label = {}]
      (ST2) at (2, 2) {ST'(3)};
      \node[node, fill = blue!50, label = {}]
      (SN) at (-2, -2) {SN(0)};
      \node[node, fill = red!50, label = {}]
      (WT) at (2, -2) {WT(2)};
      \node[node, fill = yellow!50, label = {}]
      (S) at (-3, 0) {$S$};

      \path
      (ST) edge [left] node [above] {$1-mp$} (ST2)
      (ST) edge [left] node [above] {$mp$} (WT)
      (ST2) edge [left] node [right] {$m-mp$} (WT)
      (ST2) edge [loop right] node [right] {$1-m+mp$} (ST2)

      (WT) edge [left] node [below] {$m$} (SN)
      (WT) edge [loop right] node [] {$1-m$} (WT)
      (SN) edge [left] node [left] {$1$} (S)
      (S) edge [loop right] node [] {1} (S)
      ;
    \end{tikzpicture}
    }
\caption{}
    \label{fig:nfirst}
\end{subfigure}
\begin{subfigure}{0.49\textwidth}
     \centering
       \resizebox{\textwidth}{!}{
    \begin{tikzpicture}[->,>=stealth',shorten >=1pt,auto,node
      distance=2cm,node/.style={circle,draw,inner sep=0pt,minimum size=35pt}]
      \node[node, fill = red!50, label = {}]
      (ST) at (-2, 2) {ST(3)};
      \node[node, fill = red!50, label = {}]
      (ST2) at (2, 2) {ST'(3)};
      \node[node, fill = blue!50, label = {}]
      (SN) at (-2, -2) {SN(0)};
      \node[node, fill = red!50, label = {}]
      (WT) at (2, -2) {WT(2)};
      \node[node, fill = yellow!50, label = {}]
      (S) at (-3, 0) {$S$};

      \path
      (ST) edge [left] node [right] {$m-mp$} (WT)
      (ST) edge [left] node [above] {$1-m+mp$} (ST2)
      (ST2) edge [left] node [right] {$m-mp$} (WT)
      (ST2) edge [loop right] node [right] {$1-m+mp$} (ST2)

      (WT) edge [left] node [below] {$m$} (SN)
      (WT) edge [loop right] node [] {$1-m$} (WT)
      (SN) edge [left] node [left] {$1$} (S)
      (S) edge [loop right] node [] {1} (S)
      ;
    \end{tikzpicture}
    }
    \caption{}
    \label{fig:nsecond}
\end{subfigure}
\caption{The new MC model after introducing an absorbing state $S$, when (a) the victim takes the branch, (b) the victim does not take the branch.}
\label{fig:nabsorbing}
\end{figure}

We analyse the optimal attacker strategy as in Section~\ref{sec:counter-strategy}. The model with a new absorbing state of Algorithm~\ref{algorithm:cut-off} is shown in Fig.~\ref{fig:nabsorbing}. We use SageMath to solve Inequality~(\ref{ineq:4}). However, there exist exponential polynomials of three variables $m,p,c$ and the solutions cannot be concise. We first fix the update probability $m$ to be $0.5$, which results in good performances in~\cite{JCST2021zhao}. Then we put restrictions on parameters of differential privacy to ensure that two conditional probabilities are close enough so that the attacker will not distinguish. Concretely, the restrictions of Inequality (\ref{ineq:1}) and (\ref{ineq:2}) are put forward for evert step $c$. We choose parameters $\epsilon=0.1$, $\delta = 0.01$. Then the difference of two conditional probabilities will be less than$ (e^{0.1}-1) \cdot 1  + 0.01 <0.116 $. With these values, we use SageMath to get the range of parameter $p \in [0.456,0.543]$ and choose $p=0.5$.

\begin{figure}
    \centering
    \includegraphics[width = 0.7\textwidth]{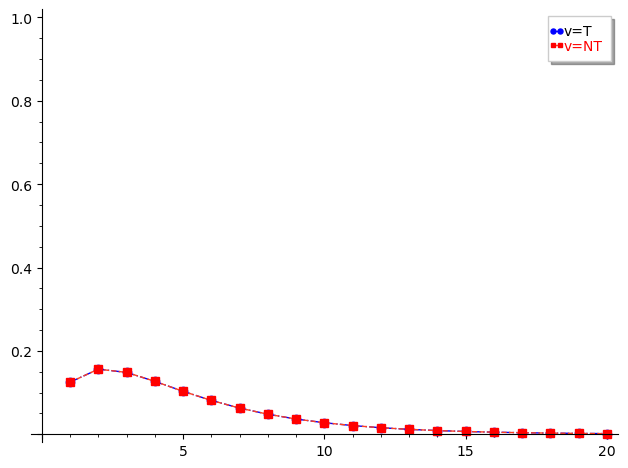}
    \caption{The line chart for the probability of observing the output $c$ when the victim takes the branch ($v=$T) or not ($v=$NT) on a new 2-bit PSC when $m=p=0.5$.}
    \label{fig:ntnt}
\end{figure}
\begin{figure}
    \centering
    \includegraphics[width = 0.7\textwidth]{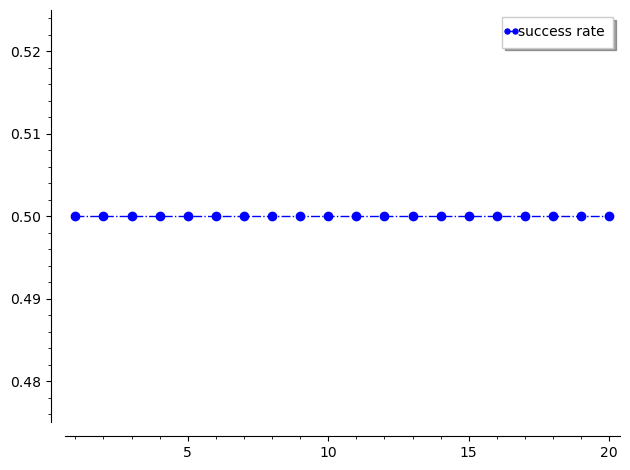}
    \caption{The line chart for the success rate of attack on a 2-bit PSC when $m=p=0.5$. }
    \label{fig:nrate}
\end{figure}

With fixed values, we can plot the probability distribution of the observed steps $c$ when the victim takes/does not take the branch respectively, and the probability of successful attack along with the steps $c$. However, we find that when $m=p=0.5$, the transitions and probabilities of the two models in Fig.~\ref{fig:nabsorbing} are the same, which indicates the conditional probabilities of observing $c$ are completely equal whether the victim takes the branch or not. In fact, the PSC satisfies $(0,0)$-differential privacy, meaning perfectly security is guaranteed. No matter which case the attacker guesses, the probability of guessing correctly is only $0.5$ and the attacker is not able to infer the execution of the victim's branch. The conditional probability of the output and probability of successful attack are shown in Fig.~\ref{fig:ntnt} and Fig.~\ref{fig:nrate}. From the perspective of defence, the new PSC is fully resistant to attack. If the design of PSCs is different or different parameters of $\epsilon,\delta$ are chosen, we can deduce different values of other parameters. However, the definition of differential privacy on PSCs still applies. We will show the actual performance under different parameters in the experiments. Before that, we theoretically calculate the misprediciton rate, which is an essential performance indicator on PSCs.

\subsection{Calculation of the Misprediction Rate}\label{sec:counter-misprediction}
In this section, we calculate the misprediction rate of PSCs theoretically. Let $\Sigma=$\{T, NT\}. The branch execution of a program is an arbitrary sequence $\Sigma^*$ and we cannot calculate the misprediction rate under all possible sequences exhaustively. In particular, for conventional deterministic saturating counters, we can even design a sequence of inputs $\sigma \in \Sigma ^*$ such that at any state the input direction is opposite to the prediction of the current state, and the misprediction rate on this sequence will be $1$. The saturating counter is completely useless in predicting in advance to speed up the reading of instructions. However, such a regular execution sequence is very rare in the practical programs, and it is not of general significance to discuss the misprediction rate on this particular sequence.

So we calculate the misprediction rate of a saturating counter from statistical views. Let the probability of taking be $s$ and the probability of not taking be $1-s$ for one specific branch. We make use of the transition matrix of the Markov chain, also called \textit{Markov Matrix}. If we prove that PSC will reach a steady state after a large number of operations on a branch, then the misprediction rate can be calculated from MC's steady state distribution. According to basic stochastic process, the row sum of the Markov matrix for each row is $1$ and whether it can reach a steady state depends on the its eigenvalues. We know that the absolute value of each eigenvalue must be less than or equal to $1$ and there must be an eigenvalue of $1$ for a Markov matrix. If there are no other eigenvalues of $1$ or $-1$, then the MC will reach the steady state. We write the Markov matrix $M$ of the PSC in Fig.~\ref{figure:nmc}, where the states are sorted in order ST, WT, WN, SN, as follows,
\hide{
\begin{align*}
    M_1 = \begin{bmatrix}
     s & t & 0 & 0\\
     s & 0 & 0 & t\\
     s & 0 & 0 & t\\
     0 & 0 & s & t
    \end{bmatrix},
   M_2 = \begin{bmatrix}
     ms+n & mt & 0 & 0\\
     ms & n & 0 & mt\\
     ms & 0 & n & mt\\
     0 & 0 & ms & mt + n
    \end{bmatrix},
\end{align*}}
\begin{align}
    M = \begin{bmatrix}
     m(qs+pt)+n & m(qt+ps) & 0 & 0\\
     ms & n & 0 & mt\\
     ms & 0 & n & mt\\
     0 & 0 & m(qs+pt) & m(qt+ps) + n
    \end{bmatrix},
\end{align}
where $m+n=1$, $s+t=1$, $p+q=1$ and $m$,$p$,$s\in [0,1]$. We use SageMath to solve the eigenvalues and it turns out that $m\neq 0$ is the sufficient condition that $M$ only exists one eigenvalue which equals to $1$. We have analyzed in the previous section that this condition is reasonable: Otherwise the saturating counter will degenerate into a static branch predictor and stay in a state forever. For other eigenvalues, unless $s$,$m$,$p$ are irrational values and satisfy particular expressions, the values cannot be $1$ or $-1$. Therefore, we can conclude that for the ordinary programs which use saturating counters for branch prediction, the counters will reach steady states after a large number of runs. Let $\mu = [a,b,c,d]$ be the steady state distribution and $a+b+c+d=1$. The steady state distribution of the matrix $M$ can be computed by solving $\mu \cdot M = \mu$. The misprediction rate can be obtained by calculating the probability of executing NT in state ST and WT, and executing T in state SN and WN, which is $r = (a + b ) \cdot t + (c + d) \cdot s$. Therefore, the steady state distribution is
\hide{
\begin{align*}
    \mu_1 = \mu_2 = \bigg[
     &\frac{s^2}{s^2+s^2t+st^2+t^2}, &\frac{s^2t}{s^2+s^2t+st^2+t^2},\\ &\frac{st^2}{s^2+s^2t+st^2+t^2}, &\frac{t^2}{s^2+s^2t+st^2+t^2}\bigg],
\end{align*}}
\begin{equation}
    \begin{aligned}
    \mu = \bigg [
    &\frac{s(qs+pt)}{s(qs+pt)(1+qt+ps)+t(qt+ps)(1+qs+pt)},\\ &\frac{s(qs+pt)(qt+ps)}{s(qs+pt)(1+qt+ps)+t(qt+ps)(1+qs+pt)}, \\ &\frac{t(qs+pt)(qt+ps)}{s(qs+pt)(1+qt+ps)+t(qt+ps)(1+qs+pt)}, \\ &\frac{t(qt+ps)}{s(qs+pt)(1+qt+ps)+t(qt+ps)(1+qs+pt)} \bigg].
\end{aligned}
\end{equation}

and the misprediction rate in steady states is
\begin{align}
\hide{
    r_1 = r_2 = \frac{st(s+2st+t)}{s^2+s^2t+st^2+t^2}, }
    r = &\frac{st(qs+pt)(1+qt+ps)+ st(qt+ps)(1+qs+pt)}{s(qs+pt)(1+qt+ps)+t(qt+ps)(1+qs+pt)}.
\end{align}

\begin{figure}
    \centering
    \includegraphics[width=0.7\textwidth]{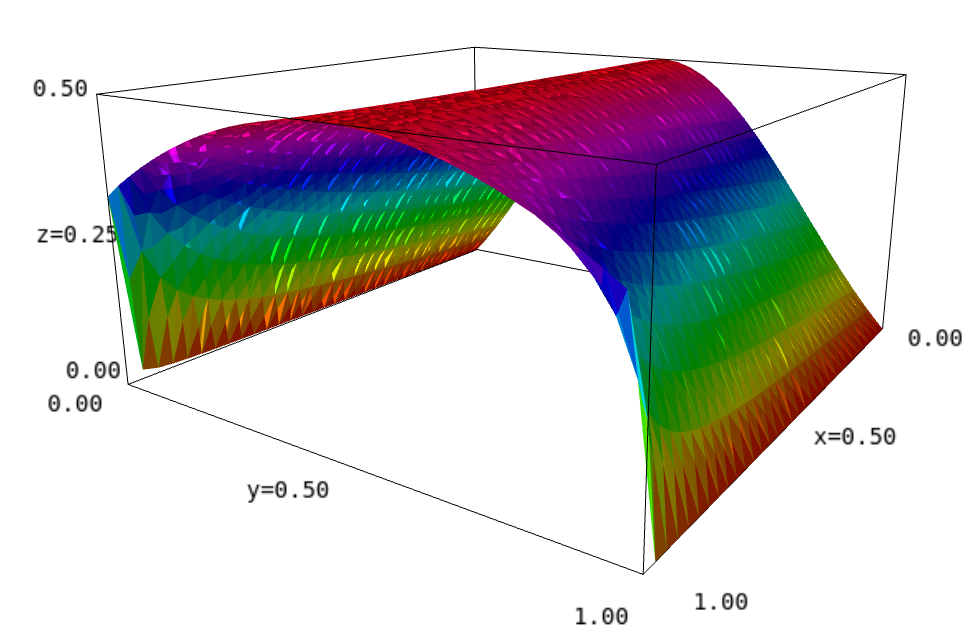}
    \caption{The misprediction rate of the new PSC in the steady state, where $x$ and $y$ represent $p$ and $s$ respectively, $p\in[0, 1]$ and $s \in [0.001,0.999]$.}
    \label{fig:misp}
\end{figure}

It can be seen that the misprediction rate of PSCs in steady states is independent of the parameter $m$. When $p=0$, the misprediction rate for the conventional saturating counter is equal to that of the original PSCs, which is only related to the probability $s$ of taking the branch, while the misprediction rate of the new PSCs is related to both parameters $s$ and $p$. Moreover, when $p=1$, $s$ cannot be $0$ or $1$. Otherwise the steady state is uncertain. We plot the misprediction rate in steady states when $p\in[0, 1]$ and $s \in[0.001,0.999]$, as shown in Fig.~\ref{fig:misp}. Note that when $p=s=0.5$, $r$ reaches the maximum value of $0.5$. At the same point, the misprediction rate of the conventional saturating counters and the original PSCs is also $0.5$. Set $p=0$ in Fig.~\ref{fig:misp}, which reflects the change of misprediction rate for conventional saturating counters and original PSCs along with the change of $s$.

\section{Experiments}\label{sec:counter-tool}
\subsection{Configurations}\label{subsec:setting}
In this section, we observe the actual performance of PSCs under different parameters on actual programs and, as a comparison, investigate the experimental results of conventional saturating counters and original PSCs. We use clock-level Gem5 simulator~\cite{binkert2011GEM5} to simulate an out-of-order execution processor, with the newest model of Intel Sunny Cove core~\cite{Wikichip} as the core of the processor. Table~\ref{tab:config} shows the experimental configurations. We implement the PSCs model on the classic branch predictor architecture Tournament~\cite{kessler1999the}. The mechanisms involving probability are implemented by \textit{Register Transfer level (RTL) code}. To evaluate performance, we adopt SPEC CPU 2017 standard processor performance test assembly~\cite{bucek2018Spec}. After being warmed up by 100 million instructions, the simulator continues to execute 100 million instructions in clock precision mode. We measure \textit{misprediction per kilometer instructions, MPKI} to demonstrate the accuracy of branch prediction and \textit{instructions per cycle, IPC} to evaluate its performance.

\begin{table}[!htbp]
    \caption{Configurations of the out-of-order processor.}
    \label{tab:config}
    \centering
    \renewcommand{\arraystretch}{1}
    \resizebox{0.7\textwidth}{!}{
    \begin{tabular}{cc}
        \hline
        Parameters & Configurations\\
        \hline
        ISA & ARM \\
        Frequency & 2.5 GHz\\
        Preocessor type & 8-decode, 8-issue, 8-commit\\
        Pipeline depth & 19 stages, fetch 4 cycles\\
        ROB/LDQ/STQ/IQ & 352/128/72/120 entries\\
        BTB & 1024 $\times$ 4-way entries\\
        PHT & Tournament: 6.3 KB\\
        ITLB/DTLB & 64/64 entries\\
        L1 ICache & 32 KB, 4-way, 64 B line\\
        L1 DCache & 48 KB, 4-way, 64 B line\\
        L2 Cache & 512 KB, 16-way, 64 B line\\
        L3 Cache & 4 MB, 32-way, 64 B line\\
        \hline
    \end{tabular}}
\end{table}

\subsection{Performance Evaluation}

\begin{figure}
    \centering
    \rotatebox{0}{\includegraphics[width=1\textwidth]{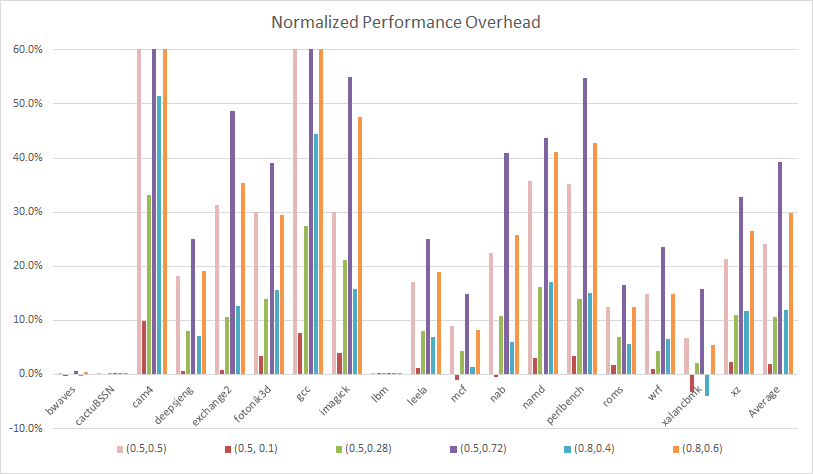}}
    \caption{The normalized performance overhead for PSCs under different settings of parameters, where baseline is the conventional saturating counter ($p=0$, $m=1$).}
    \label{fig:overhead}
\end{figure}

We measure the performance of saturating counters under different parameters as shown in Fig.~\ref{fig:overhead}. For each program, we run the instructions for 100 million times. The normalized performance overhead is the percentage of decrease in the number of the IPC compared to the conventional saturating counter ($p = 0, m = 1 $). The overhead is positive if the number of executed instructions are fewer and negative if more instructions are executed. Generally, we consider the performance overhead to be within $5\%$ is acceptable. In addition to the comparison with $m=0.5$ of the original PSC, the new PSC parameters are selected in the following way: when ($0$,$0$)-differential privacy is satisfied, select parameters $m=0.5,p=0.5$; When ($0.1$,$0.1$)-differential privacy is satisfied, choose $m=0.5$ and $m=0.8$ separately and obtain four boundary values of $p$; When ($0$,$0.2$)-differential privacy is satisfied, select parameters $m=0.5,p=0.1$. The performance and analysis are summarized as follows:

\begin{enumerate}
    \item When $m=0.5,p=0.5$, the average performance overhead is large with a value of $24.1\%$. As previously analyzed, it satisfies ($0$,$0$)-differential privacy. Although it is  impossible for the attacker to distinguish whether the victim process takes the branch, the performance for the prediction function of the PSC is greatly compromised. The reason is that the newly introduced transitions force states ST and SN to move in the opposite direction with a large probability, which seriously weakens the prediction performance of the PSC in the general programs and reduces the number of instructions processed in the clock cycle.
    \item When $m=0.5,p=0.1$, ($0$,$0.2$)-differential privacy is satisfied. The average performance overhead is the smallest with a value of $1.8\%$. Not only the secure guarantee of differential privacy is established, reflecting that the difference of two conditional probabilities is within $0.2$, but also the performance on actual programs is good.
    \item Four groups of parameters $(m,p)=(0.5,0.72)$,$(0.5,0.28)$,$(0.8,0.4)$,$(0.8,0.6)$ all satisfy ($0.1$,$0.1$)-differential privacy. But with higher values of $p$, the performance overhead get larger. Unlike the parameter $p$, which makes the state move in the opposite direction, the parameter $m$ delays updating the state of the PSC. Generally, a larger $m$ and a smaller $p$ will have a smaller impact on the prediction performance of the PSCs compared with the conventional saturating counter.
    \item Note that there are even improvements in performance on some programs with different parameters, such as $(m,p)=(0.8,0.4)$ on \textit{xalancBMK} with an improvement of $4.1\%$. This shows that on some programs, it is possible to improve the prediction performance of PSCs by appropriate delayed or reverse transitions.
\end{enumerate}

\subsection{Measure the misprediction rate in steady states}
To investigate the experimental results of the misprediction rate analyzed in Section~\ref{sec:counter-misprediction}, we refer to the experiments in~\cite{elkhouly20152bit} to evaluate that on conventional saturating counters. We apply the PSCs on the classical sorting algorithm, MergeSort (Algorithm~\ref{algorithm:merge}). The input data are integers with uniform distribution or with sorted order, respectively. Each data contains $100000$ integers. Due to the large number of inputs, each branch runs for a large number of times, which makes the PSC reach the steady state. We calculate the misprediction rate as $P_\textit{theo}$ by the probability $s$ of the branch being taken and compare it with $P_\textit{exp}$ obtained by counting the mispredictions in the experiments.

\begin{algorithm}
  \begin{algorithmic}[1]
  \Require{The integer array $list$, the left index $low$, the right index $high$}
  \Ensure{$list[low,high)$ is sorted after running the algorithm}
    \Procedure{Mergesort}{list,low,high}
    \If{low + 1 $\geq$ high}
    \Return
    \EndIf
    \State{m $\leftarrow$ $\mathrm{(high - low)/2}$}
    \State{MERGESORT(list, low, low + m)}
    \State{MERGESORT(list, low + m, high)}
    \State{leftList $\leftarrow$ list[low : low + m]}
    \Comment{Copy the list}
    \State{rightList $\leftarrow$ list[low + m : high]}
    \State{i $\leftarrow$ 0, j $\leftarrow$ 0, k $\leftarrow$ 0}
    \While {i $\neq$ leftList.size() \& j $\neq$ rightList.size()}~\label{branch1}
        \If {leftList[i] $\leq$ rightList[j]}  ~\label{branch2}
            \State{ list[low + k] $\leftarrow$ leftList[i ++]}
        \Else
            \State{list[low + k]$\leftarrow$ rightList[j ++]}
        \EndIf
        \State{k ++}
    \EndWhile
    \While {i $\neq$ leftList.size()}~\label{branch3}
        \State{ list[low + k] $\leftarrow$ leftList[i ++]}
        \State{k ++}
    \EndWhile
    \While {j $\neq$ rightList.size()}~\label{branch4}
        \State{list[low + k]$\leftarrow$ rightList[j ++]}
        \State{k ++}
    \EndWhile
    \State \Return
    \EndProcedure
  \end{algorithmic}
  \caption{MergeSort}
  \label{algorithm:merge}
\end{algorithm}

There are 4 branches in Algorithm~\ref{algorithm:merge}, namely, Line~\ref{branch1},~\ref{branch2},~\ref{branch3},~\ref{branch4}. The theoretical calculated misprediction rate $P_\textit{theo}$ and experimental results $P_\textit{exp}$ are shown in Table~\ref{tab:prob}, where SC for conventional saturating counter with $m=1$ and $p=0$. It is observed that $P_\textit{theo}$ are very close to $P_\textit{exp}$ and PSCs with $p\neq0$ usually has a larger misprediction rate than SC. The table proves the correctness of our analysis and calculation and it reminds us to choose a small $p$ in order to make the prediction accurate.
\begin{table}[!htbp]
    \caption{The experimental and theoretical misprediction rate for saturating counters on each branch in Algorithm~\ref{algorithm:merge}.}
    \label{tab:prob}
    \centering
    \resizebox{\textwidth}{!}{
\begin{tabular}{cccccc}
\hline
\multirow{2}{*}{Data} &
\multirow{2}{*}{Branch} &
  \multirow{2}{*}{Probability} &
  \multirow{2}{*}{SC} &
  \multicolumn{2}{c}{PSC} \\
 \cline{5-6}
 & & &  &
  \begin{tabular}[c]{@{}c@{}}m=0.5,\\ p=0.5\end{tabular} &
  \begin{tabular}[c]{@{}c@{}}m=0.8,\\ p=0.4\end{tabular} \\
  \hline
\multirow{8}{*}{Uniform} &
\multirow{2}{*}{\begin{tabular}[c]{@{}c@{}}Line~\ref{branch1},\\ s=0.939\end{tabular}} & $P_\textit{exp}$ & 0.061 & 0.094 &  0.089 \\
& & $P_\textit{theo}$ &0.068  & 0.114 &  0.104\\
 \cline{2-6}
& \multirow{2}{*}{\begin{tabular}[c]{@{}c@{}}Line~\ref{branch2},\\ s=0.495\end{tabular}} & $P_\textit{exp}$ & 0.510 & 0.504 &  0.506 \\
& & $P_\textit{theo}$ &0.500  &0.500  &  0.500\\
 \cline{2-6}
&\multirow{2}{*}{\begin{tabular}[c]{@{}c@{}}Line~\ref{branch3},\\ s=0.355\end{tabular}} & $P_\textit{exp}$ & 0.406 & 0.471 &  0.469 \\
& & $P_\textit{theo}$ &0.433 &0.458  &  0.453\\
 \cline{2-6}
&\multirow{2}{*}{\begin{tabular}[c]{@{}c@{}}Line~\ref{branch4},\\ s=0.437\end{tabular}} & $P_\textit{exp}$ & 0.544 & 0.514 &  0.525 \\
& & $P_\textit{theo}$ &0.487  &0.492  &  0.491\\
 \hline
 \multirow{8}{*}{Sorted} &
\multirow{2}{*}{\begin{tabular}[c]{@{}c@{}}Line~\ref{branch1},\\ s=0.891\end{tabular}} & $P_\textit{exp}$ & 0.109 & 0.154 &  0.149 \\
& & $P_\textit{theo}$ &0.128  &0.194  &  0.179\\
 \cline{2-6}
& \multirow{2}{*}{\begin{tabular}[c]{@{}c@{}}Line~\ref{branch2},\\ s=1\end{tabular}} & $P_\textit{exp}$ & 0 & 0 &  0 \\
& & $P_\textit{theo}$ &0  &0  &  0\\
 \cline{2-6}
&\multirow{2}{*}{\begin{tabular}[c]{@{}c@{}}Line~\ref{branch3},\\ s=0\end{tabular}} & $P_\textit{exp}$ & 0 & 0 &  0 \\
& & $P_\textit{theo}$ &0  &0  &  0\\
 \cline{2-6}
&\multirow{2}{*}{\begin{tabular}[c]{@{}c@{}}Line~\ref{branch4},\\ s=0.895\end{tabular}} & $P_\textit{exp}$ & 0.105 & 0.158 &  0.154 \\
& & $P_\textit{theo}$ &0.123  &0.188  &  0.173\\
 \hline
 \hide{
 \multirow{8}{*}{Descending Order} &
\multirow{2}{*}{\begin{tabular}[c]{@{}c@{}}Line~\ref{branch1},\\ s=0.895\end{tabular}} & $P_\textit{exp}$ & 0.105 & 0.158 &  0.154 \\
& & $P_\textit{theo}$ &0.123  &0.188  &  0.173\\
 \cline{2-6}
& \multirow{2}{*}{\begin{tabular}[c]{@{}c@{}}Line~\ref{branch2},\\ s=0\end{tabular}} & $P_\textit{exp}$ & 0 & 0 &  0 \\
& & $P_\textit{theo}$ &0  &0  &  0\\
 \cline{2-6}
&\multirow{2}{*}{\begin{tabular}[c]{@{}c@{}}Line~\ref{branch3},\\ s=0.891\end{tabular}} & $P_\textit{exp}$ & 0.109 & 0.155 &  0.149 \\
& & $P_\textit{theo}$ &0.128  &0.194  &  0.179\\
 \cline{2-6}
&\multirow{2}{*}{\begin{tabular}[c]{@{}c@{}}Line~\ref{branch4},\\ s=0\end{tabular}} & $P_\textit{exp}$ & 0 & 0 &  0 \\
& & $P_\textit{theo}$ &0  &0  &  0\\
 \hline}
\end{tabular}%
}
\end{table}

\section{Summary}
In this paper, we introduce the saturating counters, the basic module of branch prediction in the branch predictors, and show the attack algorithm on saturating counters. We use Markov chain to model the algorithm and analyze it. From the perspective of differential privacy, we require that the probability of the attacker observing the same output should be close enough when the victim executes the branch with taken or not-taken direction. We further study the attacker's optimal attack strategy. In order to prevent the attacker accurately guessing the execution of the victim's branch, we design a new probabilistic saturating counter, which generalizes the existing conventional and probabilistic saturating counters and prevents the attacker from accurately guessing. Moreover, considering the security guarantee of differential privacy as the constraint, we can deduce the  parameters of the saturating counters that satisfy the constraint, and theoretically calculate the misprediction rate when reaching the steady state. The simulation results show that the steady state misprediction rate of the saturating counter calculated in theory is consistent with that on the actual programs after a large number of runs. Directed by differential privacy protection, the generated parameters make the saturating counters meet the specified security requirements. Compared with the conventional saturating counters, different parameters result in different performances. When selecting suitable parameters, not only the performance is good to use on actual programs , but also the security requirement can be satisfied.

\hide{
The optimization problem is
\begin{align*}
    \min \epsilon, \\
    \mathrm{s.t.}~ \mathrm e^{-\epsilon} \mu_{\mathrm N}(n) \le \mu_{\mathrm T}(n) \le \mathrm e^\epsilon \mu_{\mathrm N}(n),  \\
    m_s,p_{s,\mathrm T},p_{s,\mathrm N} \in [0,1],
\end{align*}
}

\bibliographystyle{splncs04}
\bibliography{refs}

\end{document}

%% file: macros.tex
\newcommand{\calM}{\mathcal{M}}
\newcommand{\od}{\overline{\mathbf{d}}}

\newcommand{\calX}{\mathcal{X}}

\newcommand{\hide}[1]{}

%% file: introduction.tex
\section{Introduction}

Branch prediction is the fundamental technique to improve the instruction-level parallelism in modern high-performance processor~\cite{SamsungExynosISCA2020,IBMPower8,AMDZen2Paper}.
However, the security vulnerabilities exposed in recent years reveal that there are serious risks in modern branch predictor designs~\cite{evtyushkin2018branchscope,lee2017inferring,aciiccmez2007predicting,aciiccmez2007power,huo2020bluethunder,evtyushkin2016understanding,bhattacharya2014fault}.
Attackers exploit these vulnerabilities to obtain the branch history (i.e., taken or not-taken branches) and detect the fine-grained execution traces of process.
For example, BranchScope~\cite{evtyushkin2018branchscope} and Bluethunder~\cite{huo2020bluethunder} attacks steal secrets by detecting a specific branch predictor entry.
Furthermore, BranchShadowing~\cite{lee2017inferring} also uses the shared branch prediction histories to infer fine-grained control flow in Intel SGX.
Since the modern branch predictor resources are shared between different threads, in either a single-core processor or an SMT processor, it leaves attackers the opportunity to maliciously perceive branch history information across different processes and privileges.


The root cause of the branch predictor vulnerabilities is that the update strategy of the saturating counter is deterministic.
This makes it easy for an attacker to control the saturating counter's state to construct a side-channel attack.
As the fundamental building block of branch predictors, the saturating counter provides an excellent cost-efficient way of reducing the penalty due to conditional branches and is widely used in various branch prediction designs from simple GShare predictor to the latest TAGE-type predictor~\cite{mcfarling1993combining,Bimode1997,kessler1999the,DynamicPerceptron2001,MergingPathPerceptron2005,seznec_TAGE_SC_L,seznec_LTAGE,seznec2011newTAGE}.
However, previous studies have paid too much attention to the performance and hardware cost and ignored the security of the saturating counter.
The emergence of these security vulnerabilities reminds us to reconsider the design of this critical building block.
An promising solution is to introduce probabilistic update into saturating counter design~\cite{JCST2021zhao}. It changes the conventional deterministic state transition mode to a probabilistic state transition mode. The probabilistic saturating counter greatly reduces the ability of the attacker to perceive the saturating counter's state.
However, randomization mechanisms face challenges in security analysis.
Under-randomization compromises security, and over-randomization might cause significant performance overhead. Therefore, how to balance security and performance is an important challenge for randomization mechanisms.

Traditionally, security analysis is only through experimental methods, such analysis do not generalize sufficient theorems.
Interestingly, a prominent discipline in computer science to assure the absence of errors, or, complementarily, to find errors is formal verification.
The spectrum of key techniques in this field ranges from runtime verification, such as checking properties while executing the system, to deductive techniques such as theorem proving, to model checking. The latter is a highly automated model-based technique assessing whether a system model, that is, the possible system behavior, satisfies a security property describing the desirable behavior.

The purpose of this paper is to analyze the security of probabilistic update mechanism by combining performance evaluation with formal verification. The mechanism is modeled and analyzed in a Markov chain and the security property is specified based on the well-known privacy protection framework, differential privacy. The optimal attack strategy of side channel attack is analyzed and studied, and the computed success rate of the attack is consistent with the existing experimental data. Furthermore, analysis shows that under certain circumstances the attacker are still able to accurately guess the branch execution of the victim's process. In order to avoid always correct guessing, we design a new probabilistic saturating counter model, which generalizes the conventional deterministic and the existing probabilistic saturating counters ~\cite{JCST2021zhao}. We apply differential privacy to guide the synthesis of relevant parameters in the saturating counters in order to satisfy defense guarantees and theoretically calculate the misprediction rate when the saturating counters are in steady states. The simulation experiment shows that the calculated results of the theoretical misprediction rate of the saturating counters agree with that in actual performance. At the same time, the proper differential privacy defense guarantee can deduce useful saturating counter model parameters. Compared with the existing conventional and probabilistic saturating counters, when the parameters are selected appropriately, our newly designed models not only satisfy strict security guarantee, but also have good operation performance in actual programs.

\section{Preliminary}
A \textit{finite state machine }(FSM) is commonly used to describe a saturating counter. More specifically, we use a Moore finite state machine, or  \textit{Moore machine} in short.

\begin{definition}
  A Moore machine is defined as a tuple $M = (S, s_0, \Sigma, \Gamma, T, O)$, where
\begin{itemize}
    \item $S$ is a finite set of states;
    \item $s_0\in S$ is an initial state;
    \item $\Sigma$ is an input alphabet;
    \item $\Gamma$ is an output alphabet;
    \item $T:S\times\Sigma \xrightarrow{} S$ is a transition function, which leads the transition from the current state to the next state upon reading a symbol of the input alphabet;
    \item $O:S\xrightarrow{} \Gamma$ is an output function, mapping states to the symbols of the output alphabet.
\end{itemize}
\end{definition}

Moreover, we will adopt the standard definition of \textit{Markov chains} when probabilities are introduced into the saturating counters.

\begin{definition}
A Markov chain is a tuple $K = (S, \wp, \iota_{init})$, where
\begin{itemize}
    \item $S$ is a finite set of states;
    \item $\wp$ is a transition probability function $\wp : S \times S \xrightarrow{} [0, 1]$, for each state $s \in S$ such that
    \begin{equation*}
    \sum_{t \in S} \wp(s, t) = 1;
    \end{equation*}
    \item $ \iota_{init}: S\xrightarrow{} [0,1]$ is an initial distribution, such that$\sum_{s \in S}  \iota_{init}(s) = 1$.
\end{itemize}
\end{definition}

Our evaluation of the defense on saturating counters is based on differential privacy~\cite{D:06:DP,NRS:07:dp,DR:14:AFDP}, a privacy framework for design and analysis of data publishing mechanisms and been widely used. Let $\calX$
denote the set of \emph{data entries}. A \emph{data set} of size $n$
is an element in $\calX^n$.
Two data sets $\od, \od' \in \calX^n$
are \emph{neighbors} (written $\Delta (\od, \od') \leq 1$) if
$\od$ and $\od'$ are identical except for at most one data entry. In order to protect individual's privacy, random noise is added to the data so that the true value will not be revealed or inferred. A \emph{data
publishing mechanism} (or simply \emph{mechanism}) $\calM$ is a
randomized algorithm which takes a data set $\od$ as inputs. A
mechanism satisfies differential privacy if its output
distributions on every neighboring data set are mathematically similar.

\begin{definition}
  Let $\epsilon,\delta \geq 0$. A mechanism $\calM$ is
  \emph{$(\epsilon,\delta)$-differentially private} if for all $r \subseteq
  \textmd{range}(\calM)$ and data sets $\od, \od' \in \calX^n$ with
  $\Delta (\od, \od') \leq 1$, we have
$    \Pr (\calM (\od) \in r) \leq e^{\epsilon} \Pr (\calM (\od') \in r) + \delta.$
\end{definition}

Parameters $\epsilon$ and $\delta$ set limit to the probability differences on each neighbor for every output set $r$. Generally, the smaller values of the parameters lead to smaller differences of the probability distributions, thus better privacy protection. We will later adapt this definition to our scenario of protecting victim thread's sensitive information in the saturating counter.

\section{Saturating Counter}

\subsection{Saturating Counter}
Branch predictors are usually composed of various saturating counters which record the branch history.
A saturating counter is a Moore machine that consists of a set of states, a start state, an input alphabet, and a transition function that maps an input symbol and current state to next state. In each state, it generates an output symbol.
Fig.~\ref{fig:sc1} and Fig.~\ref{fig:sc2} are two typical saturating counter forms~\cite{ahn2009saturating}.

 \begin{figure}
 \centering
 \begin{subfigure}{0.45\textwidth}
      \centering
       \resizebox{\textwidth}{!}{
     \begin{tikzpicture}[->,>=stealth',shorten >=1pt,auto,node
       distance=8cm,node/.style={circle,draw,inner sep=0pt,minimum size=25pt}]
       \node[node, fill = red!50, label = {}]
       (ST) at (-2, 2) {ST(11)};
       \node[node, fill = red!50, label = {}]
       (WT) at (2, 2) {WT(10)};
       \node[node, fill = blue!50, label = {}]
       (WN) at (-2, -2) {WN(01)};
       \node[node, fill = blue!50, label = {}]
       (SN) at (2, -2) {SN(00)};

       \path
       (ST) edge [bend left=20] node [above] {NT} (WT)
       (ST) edge [loop left] node [] {T} (ST)
       (WT) edge [bend left=20] node [below] {T} (ST)

       (SN) edge [bend left=20] node [below] {T} (WN)
       (SN) edge [loop right] node [] {NT} (SN)
       (WN) edge [bend left=20] node [above] {NT} (SN)

       (WT) edge [bend left] node [right] {NT} (SN)
       (WN) edge [bend left] node [left] {T} (ST)
       ;
     \end{tikzpicture}
     }
 \caption{}
     \label{fig:sc1}
 \end{subfigure}
 \begin{subfigure}{0.45\textwidth}
      \centering
       \resizebox{\textwidth}{!}{
     \begin{tikzpicture}[->,>=stealth',shorten >=1pt,auto,node
       distance=8cm,node/.style={circle,draw,inner sep=0pt,minimum size=25pt}]
       \node[node, fill = red!50, label = {}, align = center]
       (T1) at (-2, 2) {T1(11)};
       \node[node, fill = red!50, label = {}, align = center]
       (T2) at (2, 2) {T1(10)};
       \node[node, fill = blue!50, label = {}, align = center]
       (T3) at (-2, -2) {T2(01)};
       \node[node, fill = blue!50, label = {}, align = center]
       (T4) at (2, -2) {T2(00)};

       \path
       (T1) edge [bend left=20] node [] {M} (T2)
       (T2) edge [bend left=20] node [] {H} (T1)
       (T1) edge [loop left] node [left] {H} (T1)
       (T2) edge [bend right=15] node [above] {M} (T3)
       (T3) edge [bend right=15] node [below] {H} (T2)

       (T4) edge [bend left=20] node [below] {H} (T3)
       (T3) edge [bend left=20] node [above] {M} (T4)
       (T4) edge [loop right] node [] {M} (T4)
       ;
     \end{tikzpicture}
     }
     \caption{}
     \label{fig:sc2}
 \end{subfigure}
 \label{fig:saturating_counter_types}
 \caption{Two typical saturating counters, for (a) predicting the branch direction, (b) selecting a branch predictor table.}
 \end{figure}

Taking the two-bit \textit{prediction counter} in Fig.~\ref{fig:sc1} as an example, it has four states: SN (strongly not taken), WN (weakly not taken), WT (weakly taken) and ST (strongly taken). There are two input symbols: T (taken) and NT (not taken), indicating the execution direction of a thread on the saturating counter.
The most significant bit of the saturating counter predicts the direction of the branch, and the other bits provides hysteresis, thereby the branch predictor requires two successive mispredictions to change the prediction of direction, which is T (taken) or NT (not taken) of the output alphabet and each is indicated by red states or blue states.
In short, a prediction counter can be used to generate taken/not-taken predictions and change its own states by reading the actual execution direction. Fig.~\ref{fig:sc2} shows another kind of saturating counter: the choice counter. Since some branch predictor architectures have several prediction tables, and a choice table is employed to pick which predictor table to use~\cite{AutomatedSatCounter}. It selects the predictor tables (T1 or T2) and changes its state when the certain table's prediction is hit (H) or missed (M).

Conventional branch predictor design allows different processes to access the same hardware resources for branch prediction directly.
The attacker process could influence predictor entries shared with the victim process to spy on the execution of sensitive branches.
Additionally, the attacker can achieve malicious training to influence the victim's (speculative) execution~\cite{evtyushkin2018branchscope,huo2020bluethunder,kocher2018spectre,chen2018sgxpectre,evtyushkin2016jump}, which, in turn, enables or exacerbates the victim's information leak.

\subsection{Prime+Probe Attack on Saturating Counter}

Branch predictor side-channel attacks require the critical capability to prime and probe saturating counters.
The saturating counter encodes the execution history information of branch instructions, which may also contain sensitive branch instructions.
The attack is the process of decoding the information in the saturating counter.
Since the state transition of a conventional saturating counter is deterministic, it is easy for an attacker to infer the direction of the victim's branch by counting the branch prediction results (correct prediction or misprediction). Take Fig.~\ref{fig:sc1} for instance. The prediction is correct if the saturating counter is in state ST or WT and the next input is T, or in state SN or WN and the next input is NT. Otherwise a misprediction occurs.
The attacker judges the prediction result of the branch predictor by measuring the execution time of probe process (e.g., reading rdtsc/rdtscp register or related hardware performance counter in x86 ISA).
In ~\cite{JCST2021zhao,evtyushkin2018branchscope}, the attacker thread tries to find the \textit{cut-off point} of the saturating counter,
which corresponds to the first time that a correct prediction shows after a bunch of mispredictions. On finding the cut-off point, the attacker can infer that the victim thread's execution direction is T or NT.

\begin{algorithm}
  \begin{algorithmic}[1]
  \Require{Prime vector $Pm=\{1,1,1,...\}$,
 the victim thread's direction $v\in\{0,1\}$ and probe vector$Pb=\{0,0,0,...\}$.}
 \Ensure{Count the number of probes before reaching the cut-off point.}
    \Function{FindingCutOff}{$Pm,v,Pb$}
      \For{$m$ in $Pm$}
        \State{execute the target branch with the direction $m$}
      \EndFor
      \State{execute the victim branch with the direction $v$}
    \State{$c \leftarrow 0$}
    \For{$b$ in $Pb$}
      \State{execute the target branch with the direction $b$}
      \If{prediction is hit}
        \Return{$c$}
      \EndIf
      \State{$c \leftarrow c+1$}
    \EndFor
    \State{\Return{$c$}}
    \EndFunction
  \end{algorithmic}
  \caption{Algorithm for finding the cut off point}
  \label{algorithm:cut-off}
\end{algorithm}

Algorithm~\ref{algorithm:cut-off} shows how to find the cut-off point and to make further analysis, we divide it into 3 phases:
\begin{itemize}
    \item Phase 1 (Line 2-3): The attacker first primes the target saturating counters to an initial
state (e.g., ST) with successive taken branches and then waits for the victim to execute.
    \item Phase 2 (Line 4): The victim thread executes its branch with taken or not-taken and changes the state of the saturating counter accordingly depending on the executed program.
    \item Phase 3 (Line 6-11):
To distinguish the execution result of the victim's branch, the probe vector used for spy must be the opposite of the prime vector used for initialization. Therefore, successive not-taken branches are executed. The attacker stops until the prediction is hit with the execution and counts the number of steps of executing the branch.
\end{itemize}

The key to distinguishing the victim's behaviors is observable differences of prediction results in the probe process measurements.
We note that there was a special point, i.e, the cut-off point, in Phase 3. Actually, it also corresponds to state SN in the saturating counter. Before this point, the attacker observed mispredictions and after this point, all are correct predictions. We counter the number of mispredictions $c$ before getting to the cut-off point, by which the victim's direction can be inferred. If the victim executes the branch with taken, the saturating counter remains in state ST and it takes $c=2$ mispredictions to reach SN; Otherwise, the saturating counter moves to state WT after the victim's execution and further it only take $c=1$ misprediction to reach SN. By repeatedly runing the algorithm, the attacker can continuously decode the victim's sensitive branch executions, which may arise in severe safety issues.